\documentclass[aps,twocolumn,showpacs,preprintnumbers,prb,superscriptaddress]{revtex4-2}

\usepackage[english]{babel}

\usepackage{graphics}
\usepackage{color}
\usepackage{amssymb}
\usepackage{amsmath}
\usepackage{overpic}
\usepackage{epstopdf}
\usepackage{threeparttable}

\usepackage{bm}
\usepackage{graphicx}
\usepackage{subfigure}
\usepackage{multirow}
\usepackage{array}

\usepackage{hyperref}
\hypersetup{plainpages=false,colorlinks=true,linkcolor=blue, citecolor=blue, urlcolor=blue}

\usepackage{longtable}
\usepackage{booktabs}
\usepackage{blindtext}

\begin{document}

\title{Lattice dynamics effects on finite-temperature stability of \textbf{\textit{R}}$_{1-x}$Fe$_{x}$ (\textbf{\textit{R}} = Y, Ce, Nd, Sm, and Dy) alloys from first principles}

\author{Guangzong Xing}
\email{XING.Guangzong@nims.go.jp}
\affiliation{Research center for Magnetic and Spintronic Materials, National Institute for Materials Science, Tsukuba, Ibaraki, 305-0047, Japan}
\affiliation{Elements Strategy Initiative Center for Magnetic Materials, National Institute for Materials Science, Tsukuba, Ibaraki, 305-0047, Japan}
\author{Takahiro Ishikawa}
\affiliation{Elements Strategy Initiative Center for Magnetic Materials, National Institute for Materials Science, Tsukuba, Ibaraki, 305-0047, Japan}

\author{Yoshio Miura}
\affiliation{Research center for Magnetic and Spintronic Materials, National Institute for Materials Science, Tsukuba, Ibaraki, 305-0047, Japan}

\author{Takashi Miyake}
\affiliation{Research Center for Computational Design of Advanced Functional Materials, National Institute of Advanced Industrial Science and Technology, Tsukuba, Ibaraki, 305-8568, Japan}
\affiliation{Elements Strategy Initiative Center for Magnetic Materials, National Institute for Materials Science, Tsukuba, Ibaraki, 305-0047, Japan}

\author{Terumasa Tadano}
\email{TADANO.Terumasa@nims.go.jp}
\affiliation{Research center for Magnetic and Spintronic Materials, National Institute for Materials Science, Tsukuba, Ibaraki, 305-0047, Japan}
\affiliation{Elements Strategy Initiative Center for Magnetic Materials, National Institute for Materials Science, Tsukuba, Ibaraki, 305-0047, Japan}
\date{\today}

\begin{abstract}
We report the effects of lattice dynamics on thermodynamic stability of binary $R_{1-x}$Fe$_x$ $(0<x<1)$ compounds ($R$: rare-earth elements, Y, Ce, Nd, Sm, and Dy) at finite temperature predicted by first-principles calculation based on density functional theory (DFT). We first demonstrate that the thermodynamic stability of $R_{1-x}$Fe$_x$ $(0<x<1)$ alloys cannot be predicted accurately by the conventional approach, where only the static DFT energy at $T = 0$ K is used. This issue can be overcome by considering the entropy contribution, including electronic and vibrational free energies, and we obtained convex hull plots at finite temperatures that successfully explain the thermodynamic stability of various known compounds. Our systematic calculation indicates that vibrational entropy helps stabilize various $R_{1-x}$Fe$_x$ compounds with increasing temperature. In particular, experimentally reported $R_2$Fe$_{17}$ compounds are predicted to become thermodynamically stable above $\sim$800 K. We also show that thermodynamic stability is rare-earth dependent and discuss its origin. Besides the experimentally reported structures, the stability of two new monoclinic $R$Fe$_{12}$ structures found by Ishikawa \textit{et al.} [Phys. Rev. Mater.~\textbf{4}, 104408 (2020)] based on a genetic algorithm are investigated. These monoclinic phases are found to be dynamically stable and have larger magnetization than the ThMn$_{12}$-type $R$Fe$_{12}$. Although they are thermodynamically unstable, the formation energies decrease significantly with increasing temperature, indicating the possibility of synthesizing these compounds at high temperatures.
\end{abstract}

\maketitle

\section{Introduction}

As one of the essential functional materials in the industry, permanent magnets (PM) have widespread applications, such as the electrical motors in vehicles, generators in the wind turbines, and hard disk drives in information storage~\cite{PM1,PM2}. The performance of the PMs can be measured by the maximum energy product ($BH$)$_{\mathrm{max}}$, which is mainly governed by two quantities, the saturation magnetization ($M_s$) and coercivity ($H_c$) that is related to the magnetocrystalline anisotropy. The challenge here is to find materials with large magnetization and strong uniaxial magnetocrystalline anisotropy.  Alloys composed of rare-earth (RE) and transition metals (TM)  are a class of PMs  satisfying these requirements and possessing the highest performance. The strong magnetocrystalline anisotropy mainly arises from the strong spin-orbit coupling of RE-4$f$ orbital shells~\cite{soc3} and the 3$d$ electrons of the TM elements can provide large $M_s$. However, RE elements are not only resource critical and expensive but also detrimental to $M_s$. As a consequence of this, TM-rich phases of RE alloys are the best candidates for high-performance PMs.

Currently, the high-performance PM in commercial use is the Nd-Fe-B-based ternary alloys, the main phase of which is Nd$_2$Fe$_{14}$B~\cite{NdFeB1, NdFeB2, NdFeB3, NdFeB4} with the magnetization and magnetocrystalline anisotropy of 1273 emu/cm$^3$ and 67 kOe~\cite{NdFeB4} at room temperature, respectively. However, the Curie temperature ($T_c$) of 585 K as measured by  Sagawa $et$ $al$ \cite{NdFeB5} is marginally above the  practical use of the high-temperature edge of 450--475 K. $T_c$ and magnetocrystalline anisotropy field can be improved through the substitution of Nd with Dy~\cite{NdFeB6}. However, Dy is considered as a less abundant element with a high price on earth. Thus, considerable efforts have been made to design or search for RE-lean or RE-free magnets~\cite{RE1,RE2,RE3,RE4,RE5}. 

In this context, two promising phases, ThMn$_{12}$-type $R$Fe$_{12}$ and $R_2$Fe$_{17}$ ($R$: rare earth) with hexagonal and rhombohedral symmetries, shown in Fig.~S1 of the SI~\cite{supplement}, have attracted renewed interest, owing to the high content of Fe with respect to the best performance magnet Nd$_2$Fe$_{14}$B. Unfortunately, although the large magnetization could be realized in the $R_2$Fe$_{17}$ class of compounds, the low $T_c$ and plane magnetic crystal anisotropy limit their application as high-performance PMs. Coey and co-workers reported~\cite{R2Fe17,R2Fe17_1} that small interstitial atoms such as N or C introduced into binary Sm$_2$Fe$_{17}$ and Y$_2$Fe$_{17}$ could elevate their $T_c$ by factor two and alter the magnetic anisotropy direction. The major problem here is $R_2$Fe$_{17}X_\delta$ ($X=$C, N, B) is thermodynamically unstable, making them easy to decompose to more stable binary phases and bcc Fe at high temperatures~\cite{R2Fe17_2,R2Fe17_3}. Besides the interstitial atoms, substitution of RE or Fe is another way to improve the performance of the $R_2$Fe$_{17}$ compounds. Both results from experiments and theoretical study prove that the substitutes from the RE site or Fe site can stabilize $R_2$Fe$_{17}X_\delta$ phase and enhance their performance. Chen and coworkers~\cite{R2Fe17_4} reported that Cr substitute helps in forming the Sm$_2$(Fe,Cr)$_{17}$ alloys with a strong uniaxial magnetocrystalline anisotropy. Sm$_2$Fe$_{17}$(C,N)$_3$ could be stabilized through substituting La or Ce for Sm without affecting the saturation magnetization too much~\cite{R2Fe17_5}. These strategies mentioned above suggest that $R_2$Fe$_{17}$ compounds are attractive candidates for permanent magnets.

When it comes to ThMn$_{12}$-type $R$Fe$_{12}$ compounds, the major obstacle with technological application is that bulk phase is difficult to be synthesized because of the thermodynamic instability, making it easy to decompose into the stable $R$Fe$_2$ phase with lower energy. Suzuki and co-workers~\cite{YFe12} reported  YFe$_{12}$ as a metastable phase obtained through the rapid quenching method. This phase is not favorable for high-performance PM as it is challenging to convert this material into highly textured magnets fully utilizing the $M_s$. Although bulk binary ThMn$_{12}$-type $R$Fe$_{12}$ compounds do not exist, a thin film phase of SmFe$_{12}$~\cite{Smfilm1, Smfilm2,Smfilm3, Smfilm4} and NdFe$_{12}$~\cite{Ndfilm1, Ndfilm2} has been successfully synthesized and widely investigated. However, a PM needs to have a certain thickness along its magnetic flux line to induce the magnetic flux outside of the material, which means thin films are not suitable as PMs~\cite{film}. Bulk phase of $R$Fe$_{12}$ can be stabilized via the substitution of a small amount of Fe with stabilizer elements, such as $M=$ Ti, Si, V, Cr and Mo to form the ternaries $R$Fe$_{12-x}M_x$~\cite{RFeM1, RFeM2, RFeM3, RFeM4, RFeM5,RFeM6, RFeM7, RFeM8, RFeM9, RFeM10, RFeM11}. The main drawback of stabilizer is that magnetization decreases fast with the increased content of $M$, which has a strong site preference and exclusively replace only one of three nonequivalent Fe sites.

Up to now, the problem of phase stability for ThMn$_{12}$-type $R$Fe$_{12}$ is not yet eliminated, especially the mechanism of forming a thin-film phase and metastable bulk phase of YFe$_{12}$~\cite{YFe12} are not clear. Formation of a thin film or metastable phase means that the driving force for phase separation is exceedingly week, and diffusion is kinetically inhibited in the nonequilibrium phase, as discussed in the context of other metastable materials~\cite{mecha1, mecha2}. This can not be elucidated when using static formation energy obtained by conventional first-principles calculation approach based on density functional theory (DFT) at $T = 0$ K, as discussed in Refs.~\cite{RFetheo1,RFetheo2,RFetheo3,RFetheo4}, of which the formation energy of $R$Fe$_{12}$ lies highly above the convex hull tie line. Motivated by this, we report in this paper the lattice dynamics effects on thermodynamic stability of the binary $R_x$Fe$_{1-x}$ $(0<x<1)$ ($R =$ Y, Ce, Nd, Sm, and Dy) compounds evaluated by considering vibrational entropy contribution at finite temperature. By going beyond the conventional prediction conducted at $T = 0$ K, we show that the formation energy in convex hull plot decreases dramatically with increasing temperature, which may explain the existence of known thin-film and metastable phases as mentioned above. This entropy-driven stabilization of various $R_{1-x}$Fe$_x$ compounds can be explained by the $x$-dependence of phonon frequencies. We also show that the formation energy is RE dependent, and the RE elements having smaller ionic radius are more effective in stabilizing the ThMn$_{12}$-type $R$Fe$_{12}$ at finite temperatures. 

The structure of this paper is organized as follows. In the next section, we describe our theoretical methods in detail. The approximation methods of calculating Helmholtz free energy and computational details are given in Secs.~\ref{subsec:free_energy} and \ref{subsec:computational_detail}. Our strategy of systematically generating the studied alloy compounds is explained in Sec.~\ref{subsec:list_of_alloys}. In Sec.~\ref{sec:results}, we show our main results. Ground state properties of promising ThMn$_{12}$-type $R$Fe$_{12}$ and $R_2$Fe$_{17}$ compounds are discussed in Sec.~\ref{subsec:ground_state}. The dynamically stable $R_{1-x}$Fe$_x$ $(0<x<1)$ compounds are identified by performing phonon calculations in Sec.~\ref{subsec:dynamical_stability}. In Sec.~\ref{subsec:thermodynamic_stability}, the thermodynamic stability is evaluated by calculating formation energy at finite temperatures. Two new types of monoclinic $R$Fe$_{12}$ are studied in Sec.~\ref{subsec:monoclinic}. We summarize this study in Sec.~\ref{sec:summary}. 

\section{Methods}

\label{sec:method}

\subsection{Helmholtz free energy}

\label{subsec:free_energy}

To evaluate the finite-temperature phase stability of the $R_{1-x}$Fe$_x$    alloy compounds in this study, we employ the Helmholtz free energy defined as 
\begin{equation}
     F(T) = E_{0}(\bm{c}_{0}) + F_{\mathrm{el}}(\bm{c}_0, T) + F_{\mathrm{vib}}(\bm{c}_0, T),
      \label{eq:free_energy}
\end{equation}
where $E_{0}(\bm{c})$ is the static internal energy of electrons obtained by a DFT calculation, $F_{\mathrm{el}}(\bm{c},T)$ is the electronic free energy, and $F_{\mathrm{vib}}(\bm{c},T)$ is the vibrational free energy at temperature $T$ and cell parameter $\bm{c}$, which is a vector comprising six parameters defining the Bravais lattice. 
In general, the thermal expansion of material is finite and thus makes $\bm{c}$ temperature-dependent, i.e., $\bm{c}(T)$, which is determined as $\bm{c}(T) = \arg \min_{\bm{c}} F(\bm{c},T)$. However, predicting $\bm{c}(T)$ for all of the studied $R_{1-x}$Fe$_x$ alloy compounds is extremely challenging because most of the studied alloys display a non-cubic lattice, for which the quasiharmonic (QH) phonon calculation needs to be conducted on a multidimensional grid of the lattice parameters. Hence, for all temperatures, we use the lattice parameters at 0 K estimated as $\bm{c}_{0}=\arg \min_{\bm{c}} E_{0}(\bm{c})$.

The electronic free energy $F_{\mathrm{el}}(T)$ is obtained based on the fixed density-of-states (DOS) approximation~\cite{Fel2} as
\begin{align}
& F_{\mathrm{el}}(T) = U_{\mathrm{el}}(T) - TS_{\mathrm{el}}(T), \label{eq:Fel} \\
& U_{\mathrm{el}}(T) =
\sum_{\bm{k}n\sigma}w_{\bm{k}}\epsilon_{\bm{k}n\sigma} [f_{\bm{k}n\sigma} -\theta(\epsilon_{\mathrm{F}}-\epsilon_{\bm{k}n\sigma})], \label{eq:Uel}\\
& S_{\mathrm{el}}(T) = - k_{\mathrm{B}}
\sum_{\bm{k}n\sigma}w_{\bm{k}}[f_{\bm{k}n\sigma}\ln{f_{\bm{k}n\sigma}} \notag \\
&\hspace{30mm}+
(1-f_{\bm{k}n\sigma})\ln{(1-f_{\bm{k}n\sigma})}]. \label{eq:Sel}
\end{align}
Here, $U_{\mathrm{el}}$ and $S_{\mathrm{el}}$ are the internal energy and the electronic entropy, respectively, and  $\epsilon_{\bm{k}n\sigma}$ represents the Kohn--Sham eigenenergy of the $n$th band with spin $\sigma$ at momentum $\bm{k}$. $f_{\bm{k}n\sigma}$ is the Fermi--Dirac distribution function defined as  $f_{\bm{k}n\sigma} = f(\epsilon_{\bm{k}n\sigma},T)= [e^{(\epsilon_{\bm{k}n\sigma}-\mu)/k_{\mathrm{B}}T}+1]^{-1}$ with $k_\mathrm{B}$ and $\mu$ being the Boltzmann constant and temperature dependent chemical potential, respectively, and $\theta(x)$ is the step function. The summation over $\bm{k}$ is performed with the irreducible set of $\bm{k}$ points in the first Brillouin zone (FBZ) with its weight $w_{\bm{k}}$. The second term on the right hand side of Eq.~(\ref{eq:Uel}) subtract $E_{0}$, which is already included in Eq.~(\ref{eq:free_energy}).
 
The vibrational free energy $F_{\mathrm{vib}}(T)$ is evaluated based on the harmonic approximation as
\begin{equation}
 F_{\mathrm{vib}}(T) = \frac{1}{N_q}\sum_{\bm{q}\nu} \bigg\{ \frac{\hbar\omega_{\bm{q}\nu}}{2} + k_{\mathrm{B}}T\ln{\big[1-\exp{(-\frac{\hbar\omega_{\bm{q}\nu}}{k_{\mathrm{B}}T})}\big]}\bigg\},
\label{eq:fvib}
\end{equation}
where $\hbar$ is the reduced Planck constant and $N_q$ is number of $\bm{q}$ points in the FBZ. The phonon frequency $\omega_{\bm{q}\nu}$ is obtained by diagonalizing the dynamical matrix 
\begin{equation}
    D_{\kappa\kappa'}^{\alpha\beta}(\bm{q})= \frac{1}{\sqrt{\mathcal{M}_{\kappa}\mathcal{M}_{\kappa'}}}
    \sum_{\ell}\Phi^{\alpha\beta}(0\kappa;\ell\kappa') e^{\mathrm{i}\bm{q}\cdot\bm{r}(\ell)}, \label{eq:dymat}
\end{equation}
where $\mathcal{M}_{\kappa}$ is the atomic mass of the $\kappa$th atom and $\Phi^{\alpha\beta}(0\kappa;\ell\kappa')$ is the second-order interatomic force constant between the $\kappa$th atom in the unit cell at the center and the $\kappa'$th atom in the $\ell$th unit cell shifted from the origin by $\bm{r}(\ell)$.

\subsection{Studied alloy compounds}

\label{subsec:list_of_alloys}

The key point of evaluating thermodynamic stability is to include as many competing phases as possible in the convex hull plot. Since the composition space, including ternary and quaternary alloy compounds, is too vast to explore by \textit{ab initio} calculations, we restrict our attention to the binary compounds in this study.

To make a comparison between our prediction and experiment, we systematically generated the competing phase structures as follows; first, experimentally-reported structure prototypes of binary $R_m$Fe$_n$ ($R=$ Y, Ce, Nd, Sm, Dy) compounds were obtained from the Inorganic Crystal Structure Database. The obtained compositions are $R$Fe$_2$, $R$Fe$_3$, $R_5$Fe$_{17}$, $R_6$Fe$_{23}$, $R$Fe$_5$, $R$Fe$_7$, $R_2$Fe$_{17}$, $R_3$Fe$_{29}$, and $R$Fe$_{12}$. Two different structures reported for $R$Fe$_3$ and $R_2$Fe$_{17}$ were also considered. By contrast, we excluded $R$Fe$_7$ and $R_5$Fe$_{17}$ because their structures are highly disordered or complicated. Next, for the selected prototypes, we substituted $R$ with Y, Ce, Nd, Sm, or Dy. In this way, we generated 45 different materials in total.

In addition to these experimentally-reported structures, we also consider two new monoclinic ($C2/m$) phases of $R$Fe$_{12}$ that have recently been reported by Ishikawa \textit{et al.} based on a structure-search genetic algorithm (GA) in Ref.~\cite{YFe12-new}. While these structures have been shown to be less stable than the ThMn$_{12}$-type $R$Fe$_{12}$ at 0 K, they are predicted to have  higher Curie temperature and larger magnetization than the ThMn$_{12}$-type $R$Fe$_{12}$, which are favorable as a high-performance PM. Since dynamical stability and finite-temperature thermodynamic stability are of vital importance for realizing these phases in the laboratory, two monoclinic ($C2/m$) structures, which are labeled as type-I and type-II following Ref.~\cite{YFe12-new}, are included in our convex hull plot. The complete list of the investigated binary $R_m$Fe$_n$ compounds is shown in Table~S1 of the SI~\cite{supplement}.

\subsection{Computational details}
\label{subsec:computational_detail}

All of the DFT calculations in this study were performed by using the projector augmented wave (PAW) method~\cite{paw}, as implemented in the Vienna \textit{ab initio} simulation package (VASP)~\cite{vasp}, within the Perdew--Burke--Ernzerhof (PBE) generalized-gradient approximation (GGA)~\cite{pbe-gga}. The $4f$ electrons for all the RE elements are frozen in the core assuming trivalent configuration. Lattice constants and atomic positions of all of the studied $R_{1-x}$Fe$_x$ compounds were carefully optimized with a kinetic-energy cutoff of 400 eV for the plane-wave expansion, and the $k$-point mesh was generated automatically in such a way that the mesh density in the reciprocal space becomes larger than 450 \AA$^{-3}$. To test the convergence with respect to the $k$-mesh density, we used a much denser mesh density of 800 \AA$^{-3}$; the difference of the total energies was less than 0.1 meV/atom, indicating convergence with the mesh density of 450 \AA$^{-3}$. For the structural optimization and phonon calculations, we used the Methfessel--Paxton smearing method~\cite{Methfessel_Paxton1989} with the width of 0.2 eV. On the other hand, the tetrahedron method with the Bl\"{o}chl correction~\cite{Blochl_PRB1994} was used for calculating $E_{0}$.

The collinear spin-polarization was applied for all the DFT calculations, including the phonon calculations. The initial local moment of 3 $\mu_B$ was set for all Fe atoms along the [001] direction. We tried both the spin moment of RE elements parallel and anti-parallel with that of Fe atoms with the corresponding initial moment of 0.3 and -0.3 $\mu_B$ and calculated the total energy of $R_{1-x}$Fe$_x$ compounds. We obtained lower energy for the anti-parallel case and used this magnetic configuration for all of the calculations. The optimized maximum, minimum and average moments of different site of Fe atoms in $R_{1-x}$Fe$_x$ compounds are listed in Table~S1 of the SI~\cite{supplement}. 

The harmonic phonon calculations were conducted by computing second-order interatomic force constants using the finite-displacement method, as implemented in \textsc{phonopy}~\cite{phonopy}. The size of the supercell for each structure was chosen so that the number of atoms became $\sim$ 100 and larger, which was sufficient to reach the convergence of $F_{\mathrm{vib}}$ within the error of 1 meV/atom at 1200 K. Here we choose the maximum temperature of 1200 K as the high-temperature limit to evaluate the entropy contribution to thermodynamic stability. This temperature is slightly above the annealing temperature used by experiments to obtain the $R_2$Fe$_{17}$ phases. The adapted supercell sizes for all compounds are summarized in Table~S1 of the SI~\cite{supplement}.

\section{Results and Discussion}

\label{sec:results}

\subsection{Ground state properties}

\label{subsec:ground_state}

\begin{threeparttable}
\centering
  \caption{Calculated lattice constants of $R_2$Fe$_{17}$ and ThMn$_{12}$-type $R$Fe$_{12}$ phases compared with experimental data at room temperature obtained from Refs.~\cite{Dy2Fe17-str,Y2Fe17-str,Sm2Fe17-str,Nd2Fe17-str,Ce2Fe17-str,YFe12,Smfilm2}. $m_\mathrm{ave}$ is the average of the calculated spin moment of Fe atoms in each alloy.}
  \label{table:cell_param}
  \small
  \begin{ruledtabular}
  \begin{tabular}{lrrrrc}
    Composition &\multicolumn{2}{c}{Calc.} &\multicolumn{2}{c}{Expt.}    &$m_\mathrm{Fe}$ \\
    \cline{2-6}
     &  $a$ (\AA) & $c$  (\AA) &  $a$ (\AA) & $c$  (\AA) & $m_\mathrm{ave}$ ($\mu_B$)   \\
    \toprule[0.15mm]
    $R_2$Fe$_{17}$-$h$ ($P$6$_3$/$mmc$) \\
    Dy & 8.413  & 8.247 & 8.453  &8.287 &2.30 \\
    Y  & 8.425  & 8.245 & 8.465  &8.298 &2.31 \\
    Sm & 8.490  & 8.283 &        &      &2.36 \\
    Nd & 8.529  & 8.315 &        &      &2.38 \\
    Ce & 8.578  & 8.362 &        &      &2.41 \\ \\
    $R_2$Fe$_{17}$-$r$ ($R\overline{3}m$) \\ 
    Dy & 8.416  &12.357 &        &      &2.30  \\
    Y  & 8.428  &12.353 &        &      &2.31 \\
    Sm & 8.517  &12.425 & 8.558 &12.447 &2.38 \\
    Nd & 8.549  &12.477 & 8.579 &12.461 &2.41 \\
    Ce & 8.584  &12.536 & 8.496 &12.415 &2.42 \\ \\
    $R$Fe$_{12}$ ($I$4/$mmm$) \\
    Dy & 8.415  &4.672  &       &       &2.20\\
    Y  & 8.418  &4.674 & 8.440\tnote{a}  &4.795\tnote{a} &2.21 \\
    Sm & 8.485  &4.676 & 8.589\tnote{b}  &4.807\tnote{b} &2.23 \\
    Nd & 8.526  &4.661 &        &       &2.22\\ 
    Ce & 8.572  &4.664 &        &       &2.23 
  \end{tabular}
  \end{ruledtabular}
  \begin{tablenotes}
  \item[a] metastable, Ref.~\cite{YFe12}
  \item[b] thin-film, Ref.~\cite{Smfilm2}
  \end{tablenotes}
\end{threeparttable}

According to the high content of Fe, $R_2$Fe$_{17}$ and ThMn$_{12}$-type $R$Fe$_{12}$ shown in Fig.~S1 of the SI are two promising compounds for high-performance PMs among the compounds studied here. We compare the calculated and experimental lattice constants of these promising compounds in Table~\ref{table:cell_param}. Here and in what follows, the calculated results are shown in the ascending order of the RE ionic radius, namely, in Dy, Y, Sm, Nd, and Ce, to highlight the RE dependence better. It is reported by experiments that $R_2$Fe$_{17}$ phase possesses two different types of crystal structures: (i) Th$_2$Zn$_{17}$-type structure with rhombohedral symmetry ($R\overline{3}m$ space group) and (ii) Th$_2$Ni$_{17}$-type structure with hexagonal symmetry ($P$6$_3/mmc$ space group). The structure of $R_2$Fe$_{17}$ phase is mainly determined by the rare earth elements. Experimentally-reported Dy$_2$Fe$_{17}$ and Y$_2$Fe$_{17}$ mainly display the hexagonal structure while the rhombohedral one is found for Sm$_2$Fe$_{17}
$, Nd$_2$Fe$_{17}$, and Ce$_2$Fe$_{17}$. We note that rhombohedral Dy$_2$Fe$_{17}$, Y$_2$Fe$_{17}$ and hexagonal Ce$_2$Fe$_{17}$ structures were also reported in Refs.~\cite{Dy2Fe17-rh-str,Y2Fe17-rh-str,Ce2Fe17-hex-str} based on different temperature treatments.

As shown in Table~\ref{table:cell_param}, the calculated values show reasonable agreement with the experimental data. For most of the studied materials, the calculation slightly underestimates the experimental lattice constants at room temperature, which can be attributed to thermal expansion as all the experimental lattice constants shown in Table~\ref{table:cell_param} were measured at room temperature. The lattice constants of ThMn$_{12}$-type $R$Fe$_{12}$ and $R_2$Fe$_{17}$ reported in Refs.~\cite{RFetheo1,RFetheo2,lattice_cons,lattice_cons1,lattice_cons2} obtained via different DFT codes are also underestimated to some extent, which are consistent with our results. The exception for cerium occurs presumably because Ce in the rhombohedral Ce$_2$Fe$_{17}$ adopts a valency of $4+$ instead of $3+$~\cite{RFetheo2}, which is quite distinct from the other RE elements. In our study, the valency of Ce in rhombohedral Ce$_2$Fe$_{17}$ is 3+. Harashima \textit{et al.}~\cite{RFetheo2} reported the lattice constants of rhombohedral Ce$_2$Fe$_{17}$ with both 3+ and 4+ valencies.  Their calculated lattice constants of Ce$_2$Fe$_{17}$ with 4+ valency ($a=8.459$ \AA{} and $c=12.513$ \AA{}) agree well with the experimental values shown in Table~\ref{table:cell_param}. Moreover, our results with 3+ valency are consistent with theirs, where the lattice constants are reported as $a=8.605$ \AA{} and $c=12.557$ \AA, indicating our results are reasonable. A slightly larger discrepancy observed for the ThMn$_{12}$-type SmFe$_{12}$ can be attributed to an interface effect or presence of TbCu$_7$ and Th$_2$Zn$_{17}$ sub-phases in the thin-film sample~\cite{Smfilm2}.

The highest and lowest magnetic moments ($m_\mathrm{Fe}$) based on different site of Fe and volume dependent magnetization ($M$) for $R_2$Fe$_{17}$ and $R$Fe$_{12}$ are shown in Table~S3 of the SI. To quantify $m_\mathrm{Fe}$ of different compounds, we also summarize the weighted average moments $m_\mathrm{ave}$ based on  all different sites of Fe shown in Table~\ref{table:cell_param}. It is clear that $m_\mathrm{ave}$ of $R_2$Fe$_{17}$ is larger that that of corresponding ThMn$_{12}$-type $R$Fe$_{12}$ compounds. However, according to a much higher concentration of Fe, $M$ of ThMn$_{12}$-type $R$Fe$_{12}$ is superior to that of $R_2$Fe$_{17}$ compounds. Moreover, the different magnetic moment ($m$[i]) of nonequivalent Fe sites in ThMn$_{12}$-type $R$Fe$_{12}$, with a trend of $m$[Fe($8i$)] $>$ $m$[Fe($8j$)] $>$ $m$[Fe($8f$)], was observed in accord with the previous calculation~\cite{soc1}.

\subsection{Dynamical stability}

\label{subsec:dynamical_stability}

\begin{figure*}[th]
 \centering
 \includegraphics[width=\textwidth]{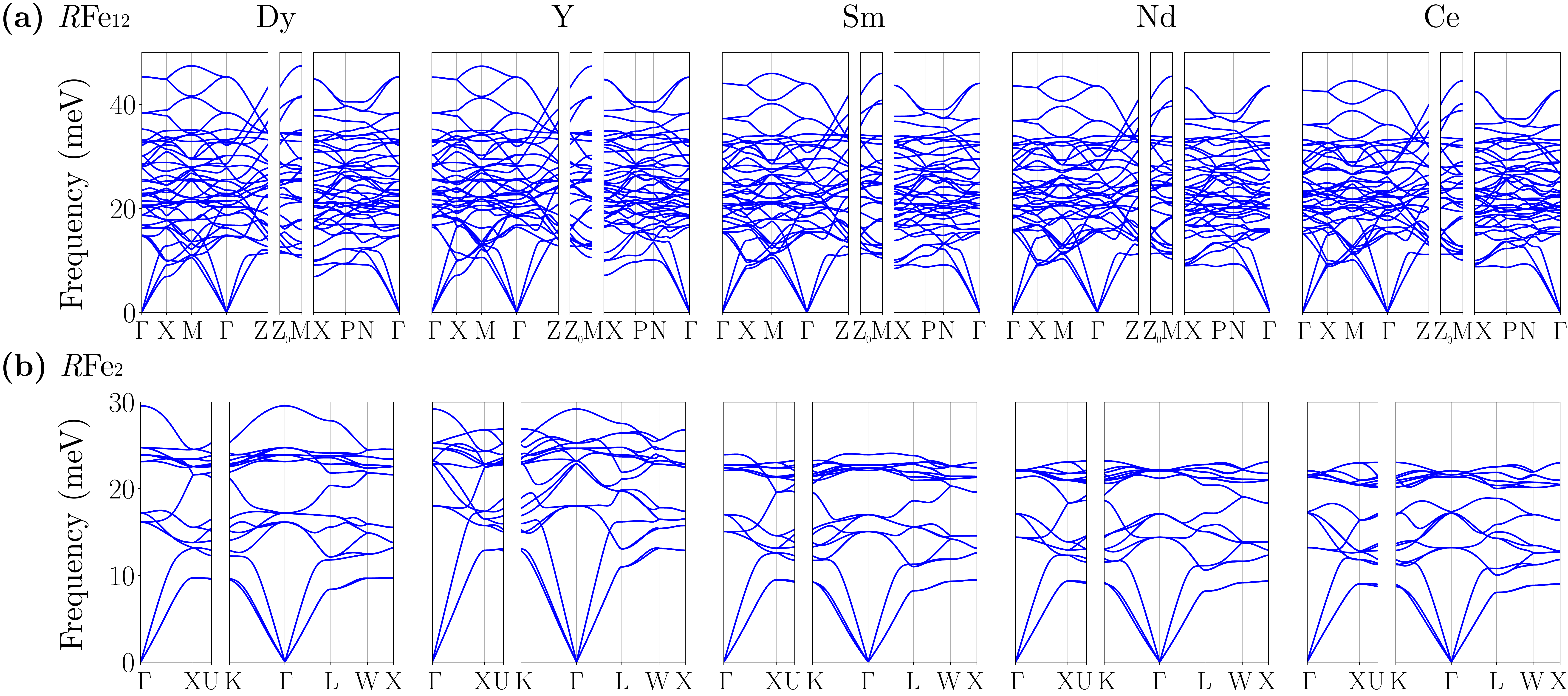}
 \caption{Calculated phonon dispersion curves of ThMn$_{12}$-type $R$Fe$_{12}$ (a) and $R$Fe$_2$ (b) compounds. The curves for $R$Fe$_{12}$ and $R$Fe$_2$ compounds with different rare earth elements of  Dy, Y, Sm, Nd, and Ce are labeled, respectively.  Note no unstable branches are present, which means all the compounds are dynamically stable. }
 \label{fig:phonon-main}
\end{figure*}

\begin{figure*}
 \centering
 \includegraphics[width=12cm]{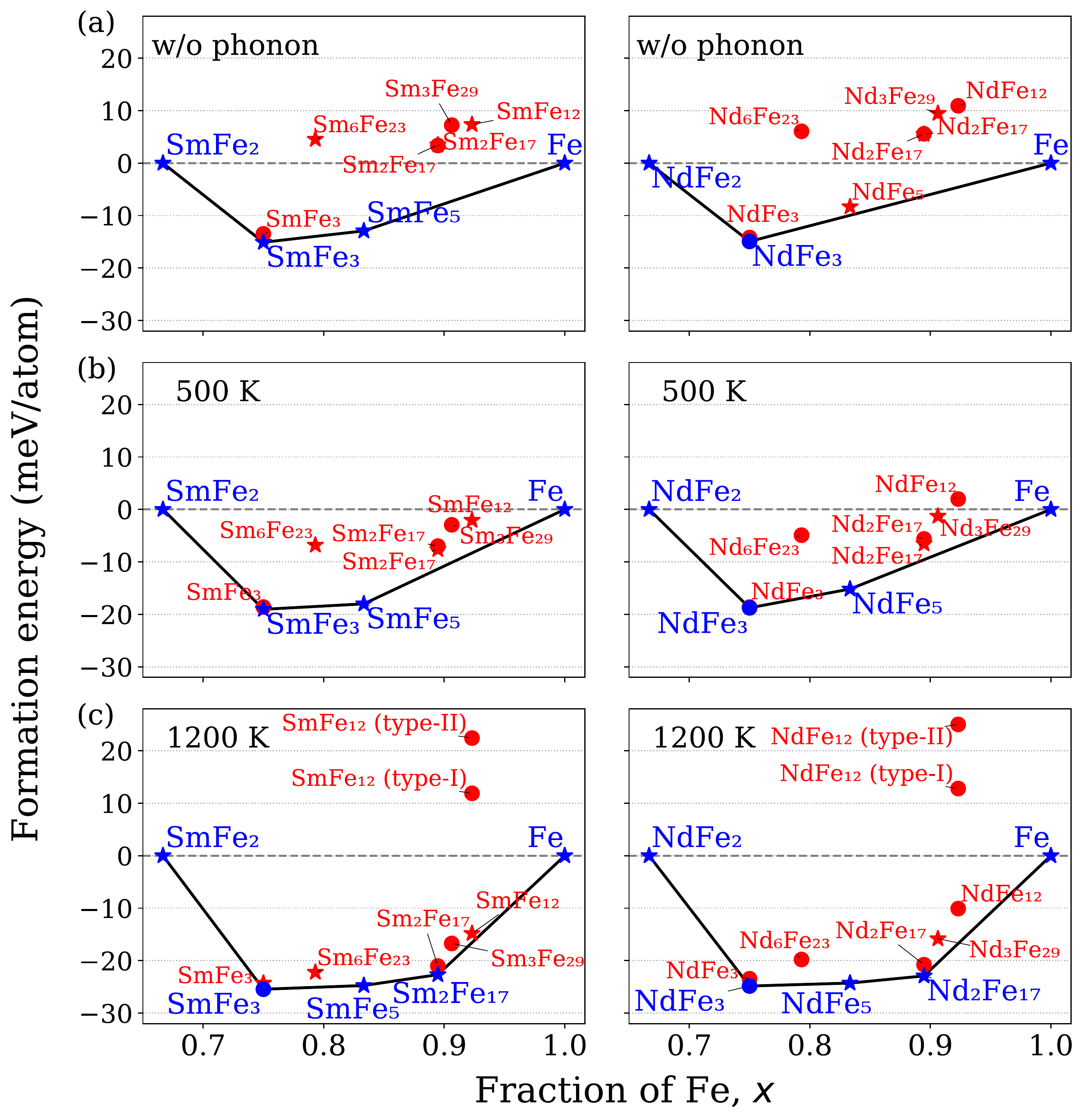}
 \caption{Change of relative formation energies per atom with respect to $R$Fe$_2$ ($R =$ Nd and Sm) and bcc Fe for different  $R_{1-x}$Fe$_x$ $(0<x<1)$ compounds obtained without vibrational free energy at 0 K (a) and with vibrational free energy at 500 K (b) and 1200 K (c), respectively. The solid line shows the convex hull and the star and circle denote the synthesized phase by experiments and hypothesized phase, respectively.  The stable and unstable phases are shown in blue and red. }
 \label{fig:convex_hull}
\end{figure*}

\begin{figure}[h]
 \centering
 \includegraphics[width=0.48\textwidth]{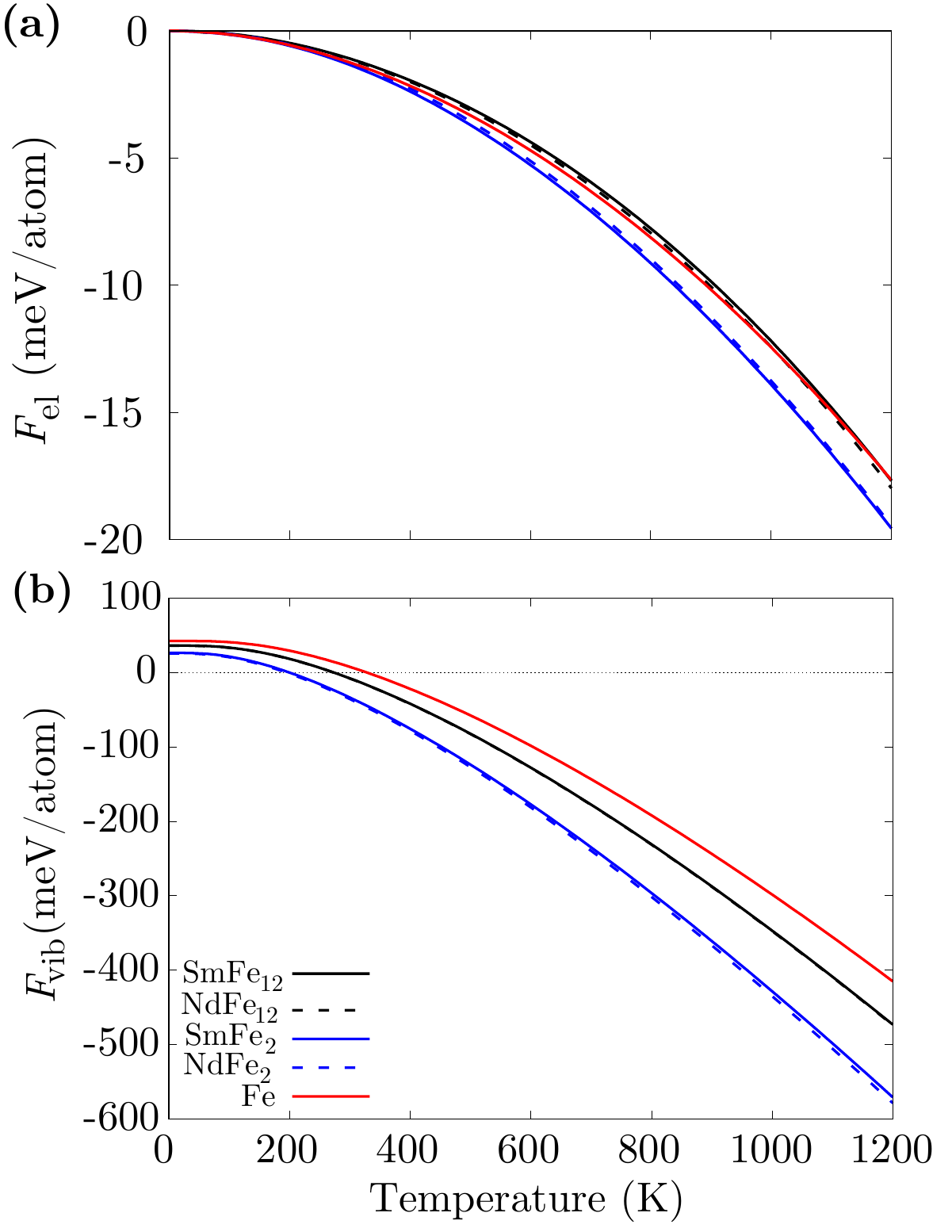}
 \caption{Calculated electronic free energy $F_{\mathrm{el}}$ (a) and vibrational free energy $F_{\mathrm{vib}}$ (b) with respect to temperature of ThMn$_{12}$-type $R$Fe$_{12}$, $R$Fe$_2$ ($R$ = Sm and Nd) phases and bcc Fe.}
 \label{fig:F_vib_el}
\end{figure}

\begin{figure}
 \centering
 \includegraphics[width=0.46\textwidth]{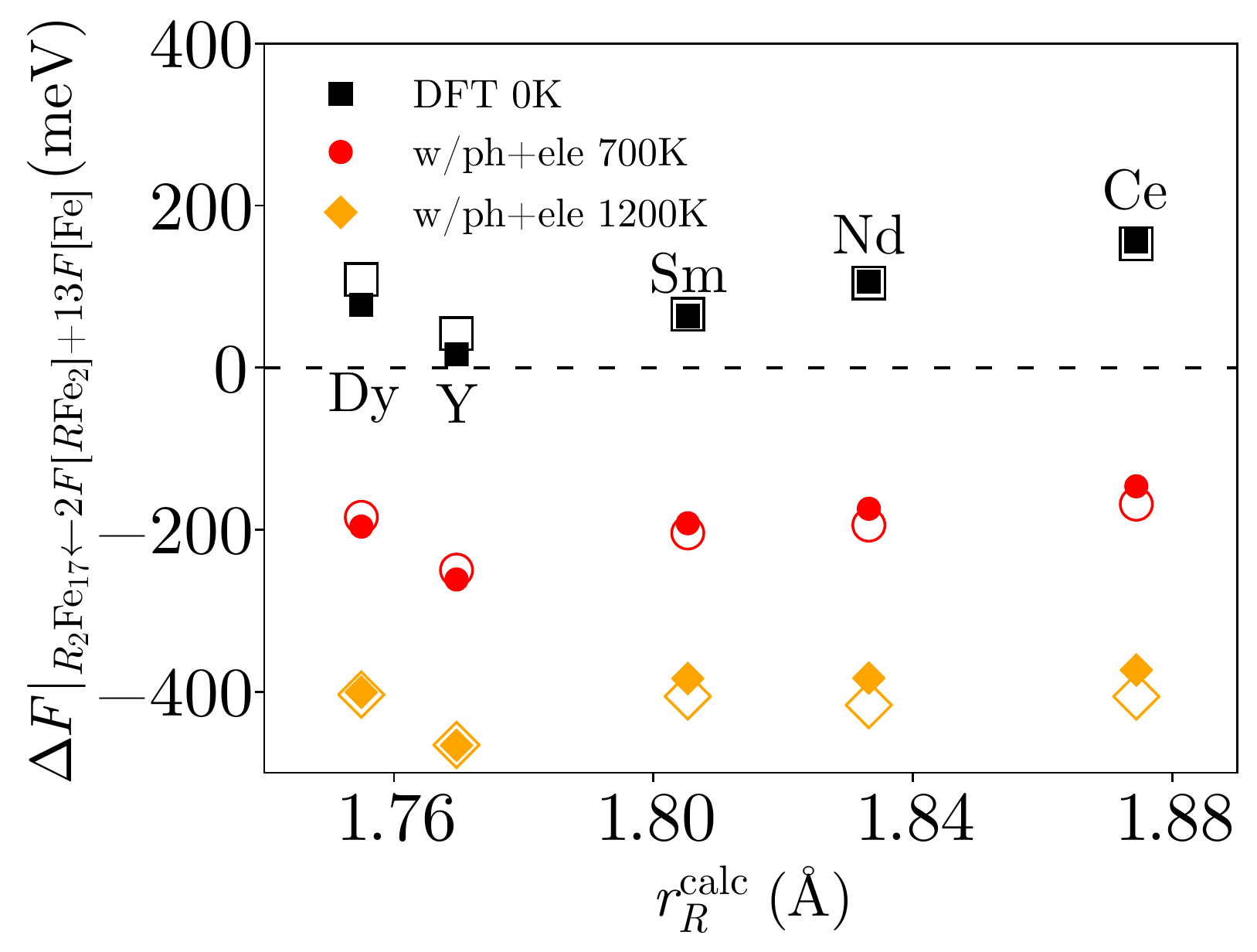}
 \caption{ Formation energy defined  for $R_2$Fe$_{17}$ ($R =$ Dy, Y, Sm, Nd, and Ce) in Eq.~(\ref{eq:formation_R2Fe17}) as a function of atomic radius at different temperatures  . The filled and open symbols denote the formation energies of $R_2$Fe$_{17}$ with hexagonal and rhombohedral structure, respectively.}
 \label{fig:R2FE17_dependent}
\end{figure}

When it comes to phase stability, the dynamical stability should be considered first, which can be evaluated by the phonon dispersion curves. At equilibrium, where the first-order force constant equals 0, a dynamically stable crystal structure means that its potential energy always increases with any small displacements of atoms. In other words, it is equivalent to the condition that no imaginary phonon modes ($\omega^2_{\bm{q}\nu}\geq 0$) are present in the phonon dispersion curves. The phonon dispersion curves for all binary compounds investigated here are listed in Fig.~\ref{fig:phonon-main} and Figs.~S2 and S3 of the SI~\cite{supplement}. As shown in these figures, no imaginary phonon modes were detected, indicating that all of the studied structures are dynamically stable at 0 K.
As evidently seen in Fig.~\ref{fig:phonon-main}(b), the highest frequency of the optical phonons gradually decreases with going from Dy to Ce, which can be attributed to the increase in the lattice constants and the associated decrease in the interatomic force constants. Moreover, YFe$_{2}$ has the largest acoustic-phonon frequencies because the atomic mass of Y is the lightest among the studied RE elements. By contrast, the RE-dependence is less obvious in $R$Fe$_{12}$ compounds (Fig.~\ref{fig:phonon-main}(a)). These significant or weak RE dependencies of the phonon frequencies characterize the RE-dependence of the vibrational free energy via Eq.~(\ref{eq:fvib}) and thereby influence the formation energy at finite temperature, as will be discussed in the following subsection.

\subsection{Thermodynamic stability}

\label{subsec:thermodynamic_stability}

The thermodynamic stability of materials can be evaluated conveniently by creating a convex hull plot. In many cases, unary compounds are selected as terminal compositions of a convex hull plot. However, for the pure RE containing $4f$ electrons, an accurate evaluation of $E_{0}$ has been challenging due to the strong correlation of localized $4f$ orbitals, for which the accuracy and convergence of DFT-GGA calculation are limited, especially when the $4f$ bands appear near the Fermi level. The open-core treatment adopted in this study helps improve the convergence, but we observed that all binary compounds $R_{1-x}$Fe$_x$ $(0<x<1)$ were predicted to be unstable against decomposition into pure RE and bcc Fe for $R=$ Sm and Nd, which does not explain the reality. This issue may be resolved by using the DFT+U method, but an optimal $U$ value would be different for different physical properties. To mitigate these technical difficulties, we use bcc Fe and $R$Fe$_{2}$ as terminal compositions in this study. The open-core treatment is appropriate to binary $R_{1-x}$Fe$_x$ compounds since $4f$ electrons do not localize near the Fermi level due to the strong hybridization between RE-$4f$ and Fe-$3d$ orbitals.

Figure \ref{fig:convex_hull} shows the convex hull plots calculated for $R =$ Sm and Nd with and without the vibrational free energy. As can be seen in the upper panel, the calculation without $F_{\mathrm{vib}}$ incorrectly predicts $R_{2}$Fe$_{17}$ phases to be unstable; the hull distance reaches over $\sim$14 meV/atom. In addition, several other reported compositions, including Sm$_6$Fe$_{23}$ and Nd$_{3}$Fe$_{29}$, are also predicted to be thermodynamically unstable. These results clearly highlight the limitation of the conventional approach based on $E_{0}$. By considering the vibrational free energy, the hull distance of these reported phases reduces gradually with increasing temperature, as shown in the middle and bottom panels of Fig.~\ref{fig:convex_hull}. For all studied RE elements, the $R_2$Fe$_{17}$ phase becomes thermodynamically stable at $\sim$ 800 K, which is reasonably close to the annealing temperature used to synthesize the main phase in the laboratory. Moreover, the hull distance of the ThMn$_{12}$-type $R$Fe$_{12}$ also decreases as the temperature increases and shows an RE dependence, which will be discussed later.

\begin{figure}
 \centering
 \includegraphics[width=0.46\textwidth]{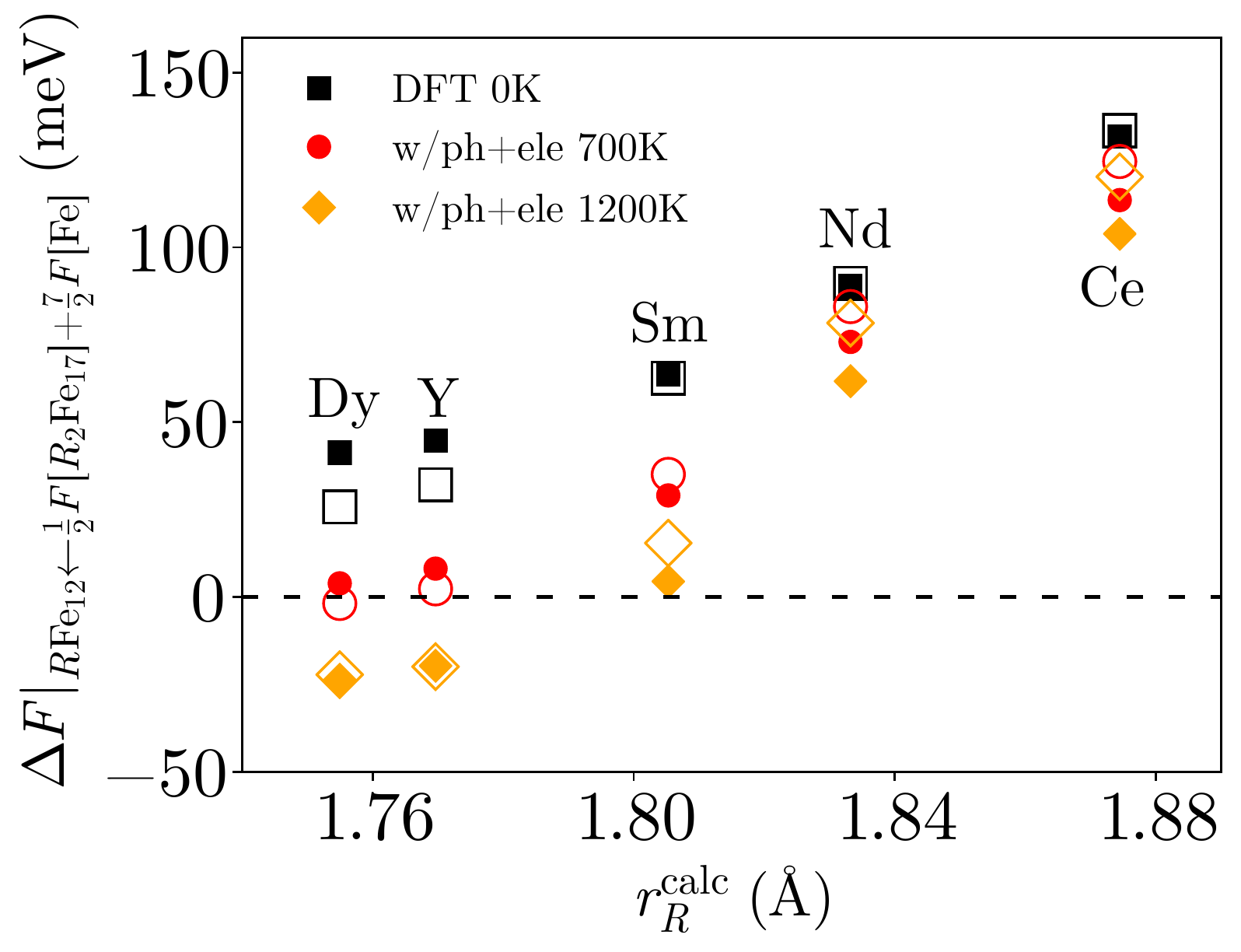}
 \caption{Formation energy defined for ThMn$_{12}$-type $R$Fe$_{12}$ ($R =$ Dy, Y, Sm, Nd, and Ce) in Eq.~(\ref{eq:RFe12}) as a function of atomic radius at different temperatures. The filled and open symbols denote the formation energies related to $R_2$Fe$_{17}$ with hexagonal and rhombohedral structure, respectively.
 }
 \label{fig:R_dependent}
\end{figure}

In the convex hull plots shown in Fig.~\ref{fig:convex_hull} and Fig.~S4 of the SI, we did not include the electronic free energy, $F_{\mathrm{el}}$, because its composition dependence is expected to be smaller than that of the vibrational free energy. To confirm this point quantitatively, we compare the calculated composition dependence of $F_{\mathrm{el}}$ and $F_{\mathrm{vib}}$ in Fig.~\ref{fig:F_vib_el} and Fig.~S5 of the SI. Due to the entropic term $-TS$, the free energy decreases monotonically with increasing temperature, at least, within the constant DOS approximation or the harmonic approximation. Hence, the composition dependence of the free energy becomes more significant in the high-temperature region. For example, we observed that the maximum difference in $F_{\mathrm{el}}$ at 1200 K, which was found between bcc Fe and $R$Fe$_{2}$, was no more than 3 meV/atom. By contrast, the maximum difference in the vibrational free energy reaches 160 meV/atom, as shown in Fig.~\ref{fig:F_vib_el}(b). These results evidently indicate the crucial importance of the vibrational free energy and the relatively minor effect of $F_{\mathrm{el}}$. The physics behind this stark contrast can be understood as follows; within the fixed DOS approximation, the temperature-dependence of $F_{\mathrm{el}}$ is manifested by the Fermi--Dirac distribution function, $f(\epsilon,T)$. The deviation of $f(\epsilon, T)$ from $f(\epsilon, 0)$, which corresponds to $f_{\bm{k}n\sigma}-\theta(\epsilon_{\mathrm{F}} - \epsilon_{\mathrm{k}n\sigma})$ in Eq.~(\ref{eq:Uel}), as well as  $f_{\bm{k}n\sigma}\ln{f_{\bm{k}n\sigma}}$ in Eq.~(\ref{eq:Sel}) are nonzero only in the range of $\pm$1 eV from $\epsilon_{\mathrm{F}}$ even at 1500 K. In this narrow energy window, the DOS is dominated by the Fe $3d$ orbitals and DOS/atom does not depend much on the fraction of Fe, $x$, leading to the weaker composition dependence of $F_{\mathrm{el}}$ per atom (Fig.~\ref{fig:F_vib_el}(a)). By contrast, since the energy scale of phonons is less than $\sim$50 meV in the studied alloys, all phonon modes contribute to $F_{\mathrm{vib}}$ significantly even below 1000 K. It can be shown easily that $F_{\mathrm{vib}}$ is an increasing function of $\omega_{\bm{q}\nu}$; thus, the negative value of $F_{\mathrm{vib}}$/atom becomes larger in its magnitude with decreasing the phonon frequency. By considering that the phonon frequency is roughly proportional to $\mathcal{M}_{\kappa}^{-1/2}$ (Eq.~(\ref{eq:dymat})) and assuming that the atomic mass can be replaced with the average mass  $\mathcal{M}_{\mathrm{avg}}(x) = (1-x) \mathcal{M}_{R} + x\mathcal{M}_{\mathrm{Fe}}$, it is easy to show that phonon frequency tends to increase (thereby $F_{\mathrm{vib}}$ tends to decrease) with increasing $x$ because $\mathcal{M}_{R} > \mathcal{M}_{\mathrm{Fe}}$ is satisfied for all the studied RE elements. This $x$ dependence of $F_{\mathrm{vib}}$/atom explains why the largest difference was observed between $R$Fe$_{2}$ ($x=\frac{2}{3}$) and Fe ($x=1$).

Interestingly, if we follow the above simple argument based on $\mathcal{M}_{\mathrm{avg}}(x)$ , it is easy to show that $F_{\mathrm{vib}}$/atom is a convex function of $x$. Indeed, we observed that the calculated data of $F_{\mathrm{vib}}$/atom roughly follows the convex curve, while some deviation from the curve was observed likely due to the $x$ dependence of interatomic force constants. The approximately convex shape of the $F_{\mathrm{vib}}$/atom curve explains qualitatively why many compounds are stabilized in the convex hull plot (Fig.~\ref{fig:convex_hull}) at finite temperatures.

Next, we discuss the RE dependence of the formation energy, particularly focusing on the Fe-rich phases, $R_2$Fe$_{17}$ and $R$Fe$_{12}$. As shown in Fig.~\ref{fig:convex_hull} and Fig.~S4 of the SI, the convex hull plots for the five different RE look somewhat similar to each other. Still, after a detailed investigation, we found a clear RE dependence. For example, the formation energy of $R_2$Fe$_{17}$ defined as 
\begin{align}
 &\Delta F|_{R_2\mathrm{Fe}_{17} \leftarrow 2 F[R\mathrm{Fe}_2]+ 13 F[\mathrm{Fe}]} \notag \\ 
 & \hspace{10mm} = F[R_2\mathrm{Fe}_{17}]-\left(2 F[R\mathrm{Fe}_2]+ 13 F[\mathrm{Fe}]\right)  \notag \\
 & \hspace{10mm}= \Delta E_{0} + \Delta F_{\mathrm{vib}} + \Delta F_{\mathrm{el}} 
\label{eq:formation_R2Fe17}
\end{align}
is shown in Fig.~\ref{fig:R2FE17_dependent} as a function of $r_{R}^{\mathrm{calc}}$, which is the radius of the RE element estimated as the half of the shortest bond length in the pure hexagonal RE phases. As already discussed in Sec.~\ref{subsec:ground_state}, $R_2$Fe$_{17}$ displays two different structures, namely, hexagonal or rhombohedral phase, depending on the RE element. Hence, the formation energy [Eq.~(\ref{eq:formation_R2Fe17})] was calculated for the hexagonal and rhombohedral phases and shown in Fig.~\ref{fig:R2FE17_dependent} by filled and open symbols, respectively. Since the free energy difference between the two structures was small, we considered $F_\mathrm{el}$ in addition to $F_\mathrm{vib}$. When we considered $E_{0}$ alone, the formation energy $\Delta E_{0}$ became positive for all the studied RE elements, and it tends to increase with increasing $r_{R}^{\mathrm{calc}}$, as shown by the square symbols in Fig.~\ref{fig:R2FE17_dependent}. With increasing temperature, $\Delta F|_{R_2\mathrm{Fe}_{17} \leftarrow 2 F[R\mathrm{Fe}_2]+ 13 F[\mathrm{Fe}]}$ changes the sign, and the RE dependence becomes weaker, which is most notable at 1200 K. The weaker RE dependence at a high temperature can be understood as follows;
first, the temperature-dependent part of Eq.~(\ref{eq:formation_R2Fe17}), i.e., $\Delta F_{\mathrm{vib}}(T)+\Delta F_{\mathrm{el}}(T)$, decreases with increasing $r_{R}^{\mathrm{calc}}$ particularly for $R=$ Sm, Nd, and Ce. This tendency was commonly observed in both hexagonal and rhombohedral phases. Since the $r_{R}^{\mathrm{calc}}$ dependence of $\Delta F_{\mathrm{vib}}(T)+\Delta F_{\mathrm{el}}(T)$ is opposite to that of $\Delta E_{0}$, the sum of these two terms shows a weaker $r_{R}^{\mathrm{calc}}$ dependence. Second, we observed for all the studied RE that the free energy difference between the hexagonal and rhombohedral structures, $F[R_2\mathrm{Fe}_{17}$-$h] - F[R_2\mathrm{Fe}_{17}$-$r]$, keeps increasing with increasing the temperature, which can be barely inferred from Fig.~\ref{fig:R2FE17_dependent} and is more clearly shown in Fig.~S6 of the SI. For Dy$_2$Fe$_{17}$ and Y$_2$Fe$_{17}$, the hexagonal phase is more stable than the rhombohedral one in the low-temperature range, while the rhombohedral phase is more stable for the other $R_2$Fe$_{17}$ compounds. With increasing the temperature, the rhombohedral phase acquires more energy gain than the hexagonal phase, thus contributing to the overall weak RE dependence of $\Delta F|_{R_2\mathrm{Fe}_{17} \leftarrow 2 F[R\mathrm{Fe}_2]+ 13 F[\mathrm{Fe}]}$ at 1200 K.

\begin{figure*}
 \centering
 \includegraphics[width=12cm]{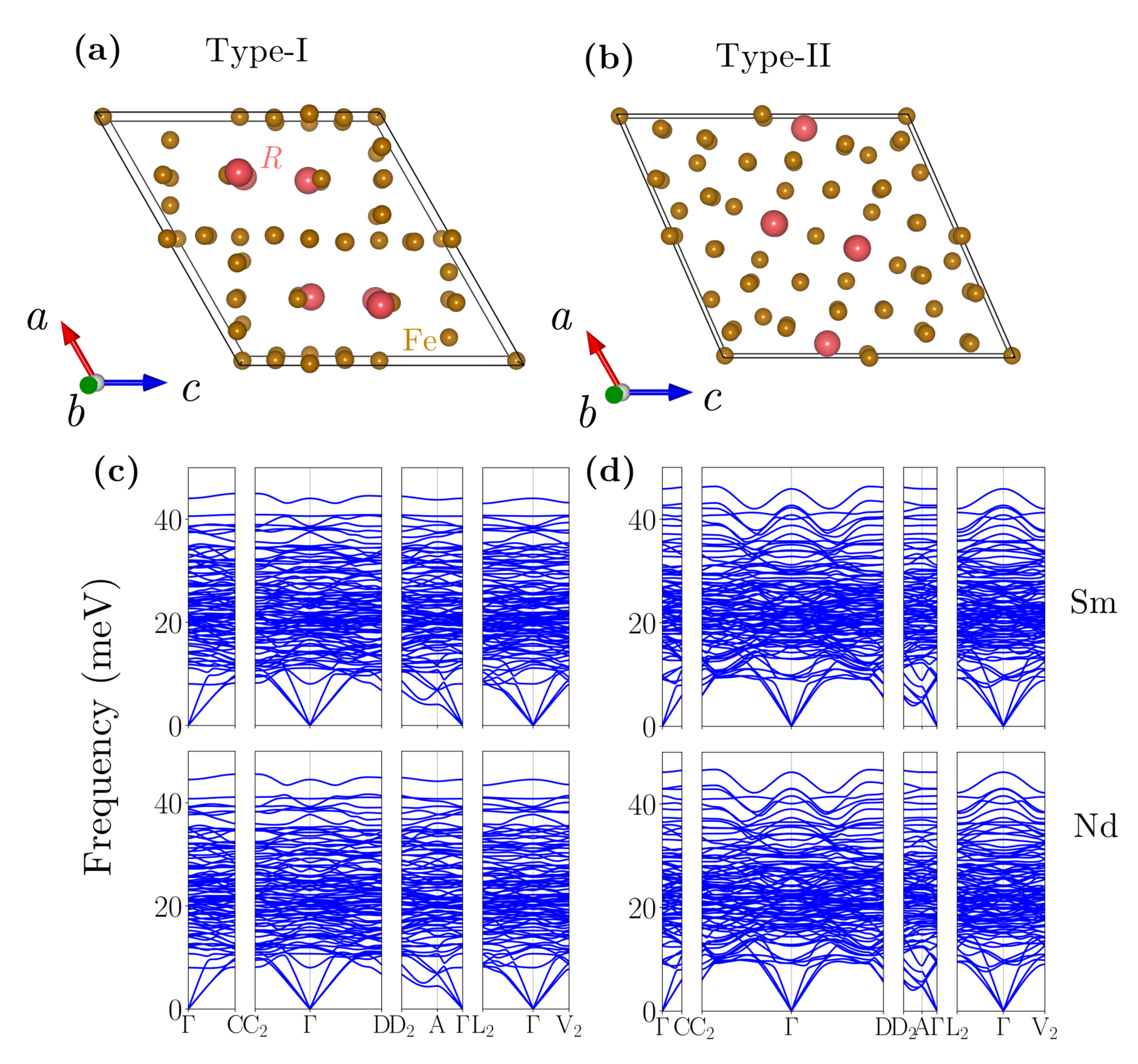}
 \caption{The low-energy $R$Fe$_{12}$ compounds: the lowest energy structure, type-I (a) and the second lowest energy structure, type-II (b) as obtained from the GA search. (c),(d) Calculated phonon dispersion curves for type-I and Type-II types of Sm- and NdFe$_{12}$, respectively. Note the predicted structures are dynamically stable. }
 \label{fig:RFe12_mono}
\end{figure*}

Finally, we discuss the thermodynamic stability of the ThMn$_{12}$-type $R$Fe$_{12}$ at finite temperature. 
A much more competitive reference phase comparing to $R$Fe$_2$ for evaluating formation energy of ThMn$_{12}$-type $R$Fe$_{12}$ compounds is $R_2$Fe$_{17}$, since all compounds of this phase studied here are realized by experiments and have a relatively lower formation energy than $R$Fe$_2$ phase. Hence, we evaluated the formation energy (meV/f.u.) by 
\begin{multline}
    \Delta F|_{R\mathrm{Fe}_{12} \leftarrow \frac{1}{2}F[R_{2}\mathrm{Fe}_{17}]+ \frac{7}{2} F[\mathrm{Fe}]} \\ 
    = F[R\mathrm{Fe}_{12}]-\left(\frac{1}{2} F[R_2\mathrm{Fe}_{17}]+ \frac{7}{2} F[\mathrm{Fe}]\right)
\label{eq:RFe12}
\end{multline}
for different ThMn$_{12}$-type $R$Fe$_{12}$ compounds with respect to the hexagonal and rhombohedral $R_2$Fe$_{17}$ phase and bcc Fe. The clear RE dependent trend is shown in Fig.~\ref{fig:R_dependent}; the smaller the atomic radius is, the more stable the ThMn$_{12}$-type $R$Fe$_{12}$ tends to be. 
Our calculation predicts DyFe$_{12}$ and YFe$_{12}$ to be thermodynamically stable above $\sim$700 K and SmFe$_{12}$ to be nearly stable when the temperature reaches 1200 K. These results are consistent with the fact that only YFe$_{12}$ has been synthesized as bulk by the high-temperature annealing followed by rapid quenching~\cite{YFe12}. Recently, Harashima \textit{et al.}~\cite{RFetheo1} reported based on DFT calculations that the hydrostatic pressure of $\sim$6 GPa reduced the formation energy of NdFe$_{12}$ and SmFe$_{12}$ by $\sim$50 meV/f.u. and $\sim$25 meV/f.u., respectively. They also reported that the pressure did not reduce the formation energy of DyFe$_{12}$.  On the other hand, the entropic stabilization at 1200 K shown in Fig.~\ref{fig:R_dependent} amounts to 70 meV/f.u., 60 meV/f.u., and 35 meV/f.u. for DyFe$_{12}$, SmFe$_{12}$, and NdFe$_{12}$, respectively. These results indicate that the temperature is more effective in stabilizing the ThMn$_{12}$-type $R$Fe$_{12}$ phase than pressure, particularly when the atomic radius is small. In addition, our result indicates that the finite-temperature stability of NdFe$_{12}$ and SmFe$_{12}$ may be enhanced by partially substituting (Nd, Sm) with (Y, Dy).

\subsection{Monoclinic \textbf{\textit{R}}Fe$_{12}$}
\label{subsec:monoclinic}

Recently, Ishikawa \textit{et al.} reported two new metastable phases of YFe$_{12}$~\cite{YFe12-new} named Type-I and Type-II with monoclinic $C$2/$m$ structures using the structure search method based on a scheme of GA~\cite{GA}. These new structures were reported to possess larger magnetization $M$ and higher Curie temperature $T_c$ than corresponding ThMn$_{12}$-type YFe$_{12}$, which is attractive as a high-performance PM material if the structure can be formed as a (meta-)stable phase. Knowing the details of the structure information, we generated four analogs of $R$Fe$_{12}$ ($R =$ Dy, Sm, Nd, and Ce) and investigated their phase stability and magnetic properties. Figure~\ref{fig:RFe12_mono}(a) and (b) shows the schematic crystal structures of the Type-I and Type-II monoclinic $R$Fe$_{12}$ compounds, whose structure parameters are listed in Table~S2 of the SI. Compared with YFe$_{12}$, the optimized lattice parameters and angles of $R$Fe$_{12}$ analogs change slightly. We also performed phonon calculations of these monoclinic compounds. As shown in Figs.~\ref{fig:RFe12_mono}(c),(d) and Fig.~S7 of the SI, all of these compounds are dynamically stable without exhibiting any unstable modes. The phonon frequency is less RE dependent than that of $R$Fe$_2$ phase.

Thermodynamic stability of these monoclinic structures at finite temperatures are evaluated and shown in Fig.~\ref{fig:convex_hull} and Fig.~S4 of the SI. We observe that all these monoclinic $R$Fe$_{12}$ compounds are thermodynamic unstable. Without the vibrational free energy contribution, the formation energies exceed more than 30 meV/atom above the $R$Fe$_2$-Fe tie line. Fortunately, they decrease dramatically as increasing temperature and become less than 15 meV/atom above tie line for all type-I $R$Fe$_{12}$ compounds at 1200 K. Although the type-I monoclinic $R$Fe$_{12}$ compounds are thermodynamically unstable, these dynamically stable phases are, at least, metastable. Thus, it is still possible to synthesize the monoclinic phase by experiments through applying a conjugated field (pressure, temperature, or surface area) as discussed in Ref.~\cite{metastable}. The magnetization $M$ of new type $R$Fe$_{12}$ compounds is shown in Table~S3. As seen, the predicted Type-I phases of $R$Fe$_{12}$ possess the largest $M$, which is consistent with that of Type-I YFe$_{12}$ compound predicted by Ishikawa $et. al.$~\cite{YFe12-new}. Given the lack of $f$ electrons in YFe$_{12}$ and high cost of DyFe$_{12}$, the new structures of SmFe$_{12}$ and NdFe$_{12}$ would be good candidates as a PM whose performance is superior to that of the corresponding ThMn$_{12}$-type compound if the phase instability problems can be solved, for example, by chemical substitution.

\section{Summary}

\label{sec:summary}

To summarize, we investigated the dynamical and thermodynamic stability of  $R_{1-x}$Fe$_x$ ($R=$ Y, Ce, Nd, Sm, and Dy) compounds by using first-principles calculations based on DFT with the effect of the vibrational and electronic entropies. By performing phonon calculations systematically, we showed that all compounds studied here are dynamically stable. We also demonstrated that the inclusion of the vibrational entropy significantly improves the prediction accuracy of the thermodynamic stability at a finite temperature compared to the conventional approach based on the static DFT energy.  

The ThMn$_{12}$-type $R$Fe$_{12}$ compounds, which are promising for PM applications, were thermodynamically unstable because of the presence of competing $R$Fe$_2$ and $R_2$Fe$_{17}$ phases. Nevertheless, the formation energies of $R$Fe$_{12}$ decreased significantly with increasing temperature, and the ThMn$_{12}$-type $R$Fe$_{12}$ phases were predicted to become stable in the high-temperature range, particularly for DyFe$_{12}$ and YFe$_{12}$ that have relatively smaller lattice constants than the other three RE cases. We showed that the observed stabilization and its RE dependence could be explained by the difference in the phonon frequencies and thereby in the vibrational free energy.
Moreover, two new metastable phases of monoclinic $R$Fe$_{12}$ predicted using a scheme of genetic algorithms were included in this study. These monoclinic phases showed larger magnetization, which is superior to that of corresponding ThMn$_{12}$-type $R$Fe$_{12}$. Although they were found to be thermodynamically unstable, the formation energies decreased dramatically with heating and became less than 15 meV/atom for the Type-I structure at 1200 K.

While our prediction that incorporates the vibrational and electronic entropies should be, in principle, more accurate than the conventional approach based on the static energy, it was still difficult to reach perfect agreements between theory and experiment. Since some other factors, including magnetic and mixing entropies, and lattice anharmonicity, are still missing in the present calculation, we expect the inclusion of these factors will improve the prediction accuracy even further, which is left for a future study.

\begin{acknowledgments}

This work was partially supported by the Ministry of Education, Culture, Sports, Science and Technology (MEXT) as the Elements Strategy Initiative Center for Magnetic Materials (ESICMM), Grant Number JPMXP0112101004 and as ``Program for Promoting Researches on the Supercomputer Fugaku'' (DPMSD).

\end{acknowledgments}

\bibliography{RFe12_cured2.bib}

\end{document}